\documentclass[floatfix,secnumarabic,amssymb,nobibnotes,nofootinbib,preprint,aps,pra, superscriptaddress,showpacs]{revtex4-2}
\usepackage{amssymb}
\usepackage[colorlinks=true, citecolor=blue, linkcolor=blue, urlcolor=blue]{hyperref}
\usepackage{times}
\usepackage{amssymb}
\usepackage{amsmath}
\usepackage{cleveref}
\usepackage[utf8]{inputenc}
\usepackage[T1]{fontenc}
\DeclareUnicodeCharacter{2212}{-}
\usepackage{mathrsfs}
\usepackage{graphicx}
\usepackage{epsfig}
\usepackage{eucal}
\usepackage{dcolumn}
\usepackage{color}
\usepackage{bm}
\usepackage{graphics,psfrag}
\usepackage{graphicx,psfrag}
\usepackage{braket}
\usepackage{subfigure}
\DeclareUnicodeCharacter{2212}{-}
\DeclareUnicodeCharacter{0393}{+}
\newcommand{\be}{\begin{equation}}
\newcommand{\ee}{\end{equation}}
\newcommand{\bea}{\begin{eqnarray}}
\newcommand{\eea}{\end{eqnarray}}
\newcommand{\ba}[1]{\begin{array}{#1}}
\newcommand{\ea}{\end{array}}

\begin{document}
\author{Sujan Maity}
\affiliation{School of Physical Sciences, Indian Association for the Cultivation of Science, 2A \&  2B Raja S. C. Mullick Road, Jadavpur, Kolkata - 700032, India}

\author{Soumik Das}

\affiliation{School of Physical Sciences, Indian Association for the Cultivation of Science, 2A \& 2B Raja S. C. Mullick Road, Jadavpur, Kolkata - 700032, India}

\author{Mainak Palit}

\affiliation{School of Physical Sciences, Indian Association for the Cultivation of Science, 2A \& 2B Raja S. C. Mullick Road, Jadavpur, Kolkata - 700032, India}

\author{Koushik Dey}
\affiliation{School of Physical Sciences, Indian Association for the Cultivation of Science, 2A \& 2B Raja S. C. Mullick Road, Jadavpur, Kolkata - 700032, India}

\author{Bikash Das}
\affiliation{School of Physical Sciences, Indian Association for the Cultivation of Science, 2A \& 2B Raja S. C. Mullick Road, Jadavpur, Kolkata - 700032, India}

\author{Tanima Kundu}
\affiliation{School of Physical Sciences, Indian Association for the Cultivation of Science, 2A \& 2B Raja S. C. Mullick Road, Jadavpur, Kolkata - 700032, India}

\author{Rahul Paramanik}
\affiliation{School of Physical Sciences, Indian Association for the Cultivation of Science, 2A \& 2B Raja S. C. Mullick Road, Jadavpur, Kolkata - 700032, India}

\author{Binoy Krishna De}
\affiliation{UGC-DAE Consortium for Scientific Research, Indore Centre, University Campus, Khandwa Road, Indore, 452001 India}
\author{ Hemant Singh Kunwar}
\affiliation{UGC-DAE Consortium for Scientific Research, Indore Centre, University Campus, Khandwa Road, Indore, 452001 India}

\author{Subhadeep Datta}
\affiliation{School of Physical Sciences, Indian Association for the Cultivation of Science, 2A \& 2B Raja S. C. Mullick Road, Jadavpur, Kolkata - 700032, India}
\email{sspsdd@iacs.res.in}

\title{Electron-Magnon Coupling Mediated Magnetotransport in Antiferromagnetic van der Waals Heterostructure}

\begin{abstract}
\bfseries{Electron-magnon coupling reveals key insights into the interfacial properties between non-magnetic metals and magnetic insulators, influencing charge transport and spin dynamics. Here, we present temperature-dependent Raman spectroscopy and magneto-transport measurements of few-layer graphene (FLG)/antiferromagnetic FePS\(_3\) heterostructures. The magnon mode in FePS\(_3\) softens below 40 K, and effective magnon stiffness decreases with cooling. Magnetotransport measurements show that FLG exhibits negative magnetoresistance (MR) in the heterostructure at low fields (\(\pm 0.2 \, \text{T}\)), persisting up to 100 K; beyond this, MR transitions to positive. Notably, as layer thickness decreases, the coupling strength at the interface reduces, leading to a suppression of negative MR. Additionally, magnetodielectric measurements in the FLG/FePS\(_3\)/FLG heterostructure show an upturn at temperatures significantly below ($T_\text{N}$), suggesting a role for the magnon mode in capacitance, as indicated by hybridization between magnon and phonon bands in pristine FePS\(_3\) \textit{via} magnetoelastic coupling.}

\end{abstract}

\maketitle

\section{Introduction}

The interaction between conduction electrons and spin fluctuations in magnetic insulators, \textit{i.e.}, magnons, may cause electronic and/or magnonic spin current which may be probed either electronically, by measuring spin-mixing conductance, or thermally, by spin
Seebeck effect (SSE)  \cite{RevModPhys.76.323, PhysRevLett.81.705, PhysRevLett.97.187201, PhysRevB.78.174404, PhysRevB.89.235439, PhysRevB.108.144420}. Theoretically, scattering of electron and magnon (frequency $\omega$) results in transition from electronic state $i$ to $j$ with momentum transfer ($\mathbf{q}$), and can be expressed in terms of electron-magnon coupling (EMC) matrix element ($\left\langle \Psi_{j,\mathbf{k+q}, \downarrow} \left| \nabla^{\nu}_{\mathbf{q}}V\right| \Psi_{i,\mathbf{k}, \uparrow} \right\rangle \omega_{\mathbf{q}\nu}^{-1/2}$), where $V$ represents the electron-magnon self energy, analogous to the electron-phonon and electron-photon coupling \cite{PhysRevLett.93.185503,attaccalite2010doped,PhysRevB.106.205303}. However, this process requires a spin flip in the electronic state to satisfy the angular momentum conservation, whereas electron-phonon coupling operates independent of spin state. Depending on the EMC strength, efficient momentum transfer at the metal/antiferromagnetic insulator (AFMI) interface increases the effective local density of state (DOS), which in turn, influences the electron transport through the metallic channel. Moreover, as the lattice modulation in AFMIs is accompanied by magnetic order, magnons may couple with the electric dipoles which may be captured in magnetocapacitance (MC) measurement \cite{nature2003, PhysRevB.103.064431, adma.200500737}.

Effectiveness of magnon (frequencies GHz to THz, lifetime microseconds to nanoseconds) detection over a large distance (several micrometers) \textit{via} magnetotransport/spin transport method rely on: (i) the interface between the magnetic material and the metal layers, (ii) the shape and size of the magnetic material, (iii) availability of low-temperature for the reduction of scattering with phonons, and (iv) the frequency range of the detection system \cite{PhysRevX.9.011026}. Even though AFM magnons (THz range frequencies) are gaining interest due to faster dynamics, and robustness against external fields, the detection technique is challenging compared to ferromagnets because of the lack of net magnetization, which makes conventional magneto-optical techniques less effective \cite{jungfleisch2018perspectives}. Detection of spin Hall magnetoresistance (SMR) in AFMIs is more challenging than FMIs as the coupling between the spin current and the complex magnonic structure of the AFM produces subtle changes in spin-mixing conductance than that of ferromagnets even at cryogenic temperature  \cite{10.1063/1.4997588, ramos2021ultra}. In an electron-doped antiferromagnet, spiral phase of the sublattice magnetization is not energetically favorable due to the presence of homogeneous phase. Even though magnon exchange can bind two electrons, it is weaker at large distances ($r$) between electron, decay faster as $\sim \frac{1}{r^{4}}$ than in holes ($\sim \frac{1}{r^{2}}$) \cite{PhysRevB.75.214405}. 
\newline
Consequently, for the detection of the electron-magnon coupling, graphene as transport channel with low spin-orbit coupling, high spin diffusion length ($\sim$ $\mu$m at room temperature), high carrier concentration ($\sim$ 10$^{13}$ cm$^{-2}$) is preferable over metallic films \cite{han2014graphene, RevModPhys.92.021003}. To probe magnon-mediated fluctuation in carrier density of graphene, an uncompensated interface, where the electrons on the graphene surface couple asymmetrically to the AFM sublattice can increase the EMC \cite{PhysRevB.100.100503, PhysRevB.104.125125}. An effective approach for generating a magnon spin current involves using charge-spin conversion through spin-orbit coupling, along with continuous spin transfer between electrons and magnons \cite{cornelissen2015long, PhysRevLett.118.147202, PhysRevB.86.214424}, a process referred to here as charge-magnon conversion. This conversion efficiency is much higher for uncompensated interface and the efficiency is enhanced with the thickness of AFM in two dimensional regime as discussed by Liao \textit{et. al.} \cite{PhysRevB.102.115152}. Recently, Yang \textit{et.al.} studied graphene/van der Waals AFM heterostructure (HS) and observed an unconventional manifestation of the quantum Hall effect which can be due to the presence of counterflowing spin-polarized edge channels originating from the spin-dependent exchange shift in graphene \cite{yang2024electrostatically}. Moreover, in the stacked HS (graphene/vdW FM CrBr$_{3}$/graphene), tunneling magnetoresistance (TMR) at low temperature indicates magnon emission with a possibility of spin-injection \cite{ghazaryan2018magnon}. Practically, for 2D AFMIs, thicker flakes (thickness $\geq$ 5 nm), with higher magnetic moment and  magnetic ordering temperature, may be transferred onto the graphene channel for stronger EMC \cite{PhysRevMaterials.7.045201}. In this context, few-layer graphene (FLG), compared to monolayer, provides better signal-to-noise ratio, higher sensitivity due to the increased carrier density, and greater robustness against the strain/rippling effect during micromanipulation process \cite{PhysRevLett.97.016801}. An alternative approach towards realization of AFM magnon-coupled electronic device could be correlating spin-charge-lattice coupling \textit{via} Raman scattering with conventional magnetoresistance (MR) and magnetocapacitance (MC) measurement of  FLG/vdW antiferromagnet HS in suitable geometry.

\begin{figure*}[ht]
\centerline{\includegraphics[scale=0.4, clip]{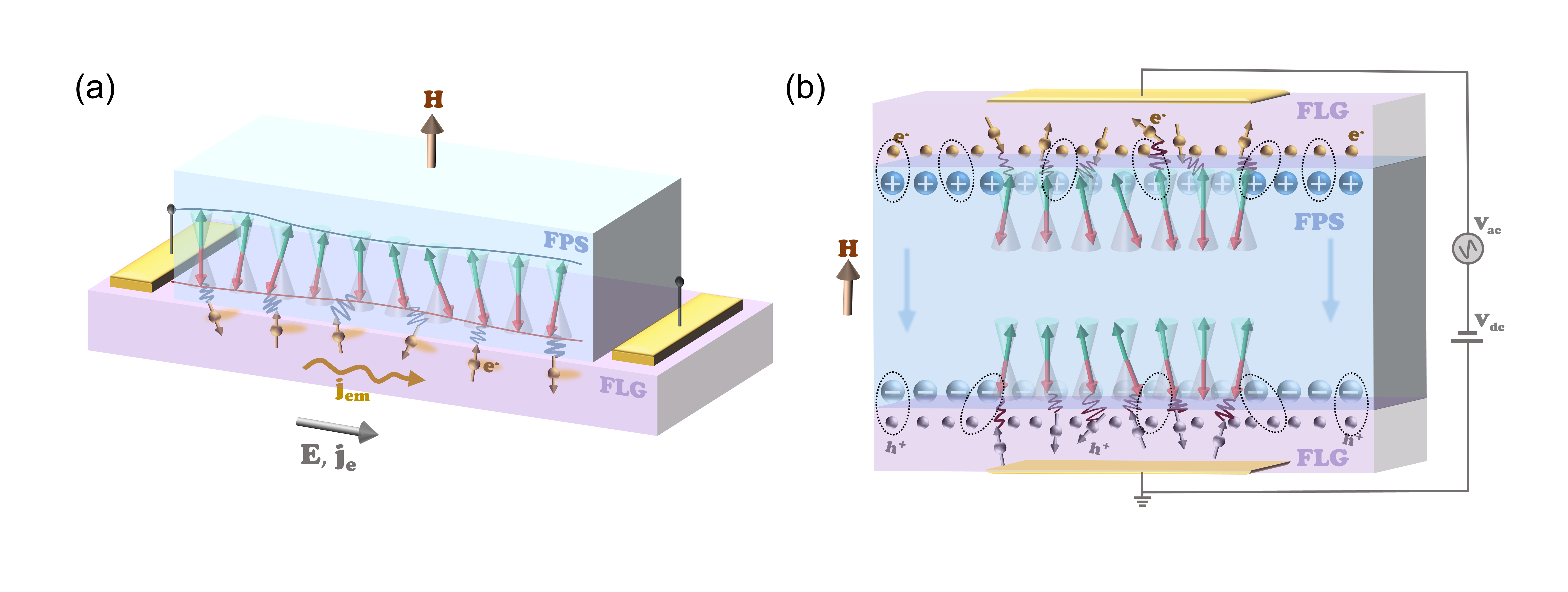}}
\caption{{(a) Schematic representation of electron-drag mediated by magnons in the FLG-FePS$_3$ heterostructure in the presence of an out-of-plane magnetic field ($H$). An electric field ($E$) is applied to the FLG layer to produce an electric current ($j$) along its direction. The itinerant electrons in the FLG layer interact with the antiferromagnetic magnons at the interface, resulting in electron-magnon coupling via electrostatic interaction. Momentum transfer from the magnons to the electrons assists in their movement along the direction of the electric field, leading to an additional current ($j_{\text{em}}$). (b) Schematic representation of out-of-plane capacitance measurement in the FLG-FePS$_3$-FLG device in the presence of a dc bias ($V_{\text{dc}}$) and magnetic field ($H$). Interfacial polarization develops between the surface-terminated charges of FePS$_3$ and the carriers in the FLG at the interfaces, as indicated by the dotted ellipses. The electrons, coupled with magnons, are represented with a spin, as the scattering event flips the spin states to conserve total angular momentum. The electron-magnon coupling influences the interfacial polarization and capacitance.
}\label{fig1}} 
\end{figure*}

Here, we bring together temperature dependent magnetotransport study (MR, and MC) and Raman  spectroscopy in a FLG/bulk 2D AFMI FePS$_{3}$ (out-of-plane anisotropy with $T_\text{N}$ $\sim$ 120 K) HS device. Similar to the pristine FePS$_3$, spin-phonon coupled modes in the HS, indicative of magnetic ordering, were observed at/around 120 K, but with 3 times higher deviation (softening) from usual anharmonicity ($\Delta\omega$) when placed on FLG. Moreover, in the HS, contrary to the characteristics of magnetic Bragg peaks, magnon mode of FePS$_{3}$ ($\sim$ 120 cm$^{-1}$) softens between 5 K and 120 K which suggests decrease of effective magnon stiffness in presence of graphene. In the HS device with micro-electrodes only on FLG, besides the typical positive MR (in isolated FLG) resulting from the distortion of the trajectory of charge carriers with increasing magnetic field, negative slope has been recorded at low magnetic field ($\leq$ 0.2 T) persisting upto 100 K which is evocative of the EMC due magnon-to-charge conversion (see Figure \ref{fig1}(a)). Predictably, with lowering the thickness of AFMI, suppression of said negative MR in FLG has been observed due to the lessened magnetic moment in the few-layer limit of FePS$_3$ which causes considerable drop in magnon-to-charge conversion and hence, reduced EMC. Further, MC of bulk FePS$_{3}$, sandwiched between two FLG flakes, exhibits an upturn below magnon temperature ($\sim$ \text{90} K) which suggests the role of magnon-phonon coupling in FePS$_{3}$ on the interfacial electric polarization (see Figure \ref{fig1}(b)). The material perspective, and the device architecture reported here may have significant implications in future magnetoelectric memory and logic devices based on antiferromagnetic magnon.

\section{Experimental Details}
Bulk Graphite crystals were purchased from NGS Naturgraphit GmbH. Single crystal FePS$_{3}$ was grown by chemical vapor transport method as described in ref. \cite{maity2024manipulating}. Two types of HSs comprised of FLG and FePS$_{3}$, both exfoliated from its bulk counterpart, were created, and denoted as (i) HS-1: FLG ($\sim$ 9 nm)/FePS$_{3}$($\sim$ 95 nm), and (ii) HS-2 : FLG ($\sim$ 5 nm )/FePS$_{3}$ ($\sim$ 30 nm) in the rest of the paper (see supplementary information for details) \citep{maity2024manipulating}.  Thickness of the individual flakes of heterostructures were confirmed by atomic force microscopy (Asylum Research MFP-3D). The average thickness of FLG flakes are $\leqslant$ 10 nm. Temperature-dependent Raman measurements  were carried out by using Jobin Yvon Horibra LABRAM-HR 800 visible micro Raman system and a 473 nm wavelength radiation from a diode laser as an excitation source. All measurements were performed under high vacuum (10$^{-6}$ mbar) using a liquid helium cryostat (Janis ST-500). Laser beam was focused through a microscope objective with 50x magnification and a spot size of $1\mu$m. Laser power was kept below $ 200 \mu$W  to avoid sample heating. For device fabrication, electrodes were patterned by e-beam lithography (Zeiss Sigma 300 with Raith Elphy Quantum) followed by metallization of Ti/Au (10 nm/70 nm). The magnetotransport measurements of the devices were carried out in a closed cycle cryostat (Advanced Research Systems) adapted to an electromagnet of $\pm$ 1 T (Holmarc Opto-Mechatronics Ltd.). Further, using the same cryostat-magnet arrangement, capacitance \textit{vs.} voltage (C-V) measurement at constant magnetic field  were carried out on FLG/FePS$_{3}$/FLG device (top-bottom contact with gold) using impedance analyzer (MFIA, Zurich Instruments) (see supplementary information for details) and temperature controller (CRYOCON 22C). Capacitance have been measured on this heterostructure by superimposing \text{500} mV peak-to-peak ac signal (frequency = \text{100} kHz) with a dc voltage varying up to $\pm$ \text{2} V. At each temperature, magnetic field have been varied up to $\pm$ \text{0.8} T \textit{via} an electromagnet along the applied voltage direction using a closed cycle cryostat and gold contacts have been fixed to the extended FLG from both the surfaces of FePS$_{3}$.

\section{RESULTS AND DISCUSSIONS}

\begin{figure*}[ht]
\centerline{\includegraphics[scale=0.5, clip]{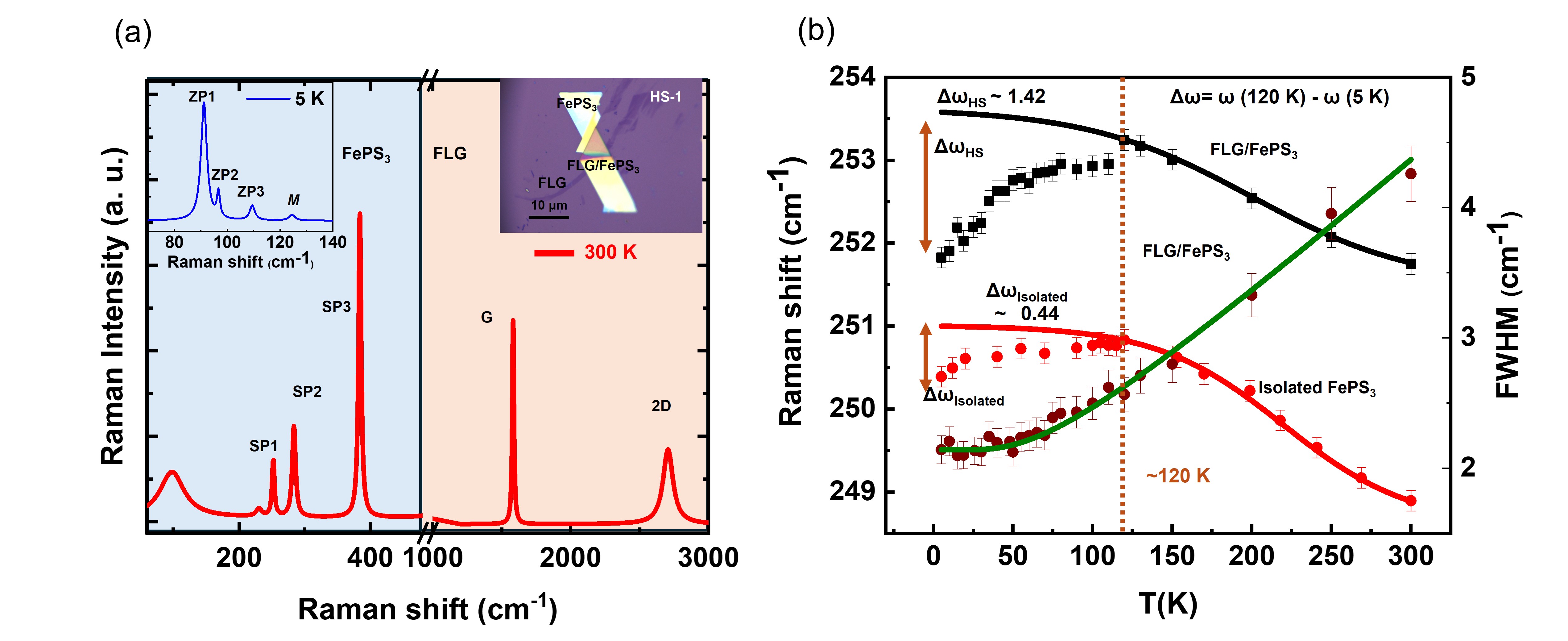}}
\caption{{Raman spectroscopy of (a) FLG/FePS$_{3}$ heterostructure (HS-1) at room temperature, obtained with an excitation wavelength of 473 nm. The spin-phonon coupled peaks are labeled as SP (1-3). The emergence of distinct phonon modes from both FLG and FePS$_{3}$ in the FLG/FePS$_{3}$ heterostructure indicates the formation of a heterointerface. The inset shows the appearance of three zero-energy phonon (ZP) modes and a magnon mode ($M$) at low temperature, which are not observed at room temperature. (b) The temperature-dependent spin-phonon coupled mode ($\sim$ 250 cm$^{-1}$) is plotted for isolated FePS$_{3}$ and the FLG/FePS$_{3}$ heterostructure (HS-1). The characteristic N\'{e}el temperature remains invariant ($\sim$ 120 K) for both samples. The deviation from anharmonicity, $\Delta\omega$, increases by more than three times in the heterostructure compared to the isolated flake.} \label{fig2}} 
\end{figure*}

\begin{figure*}[ht]
\centerline{\includegraphics[scale=0.5, clip]{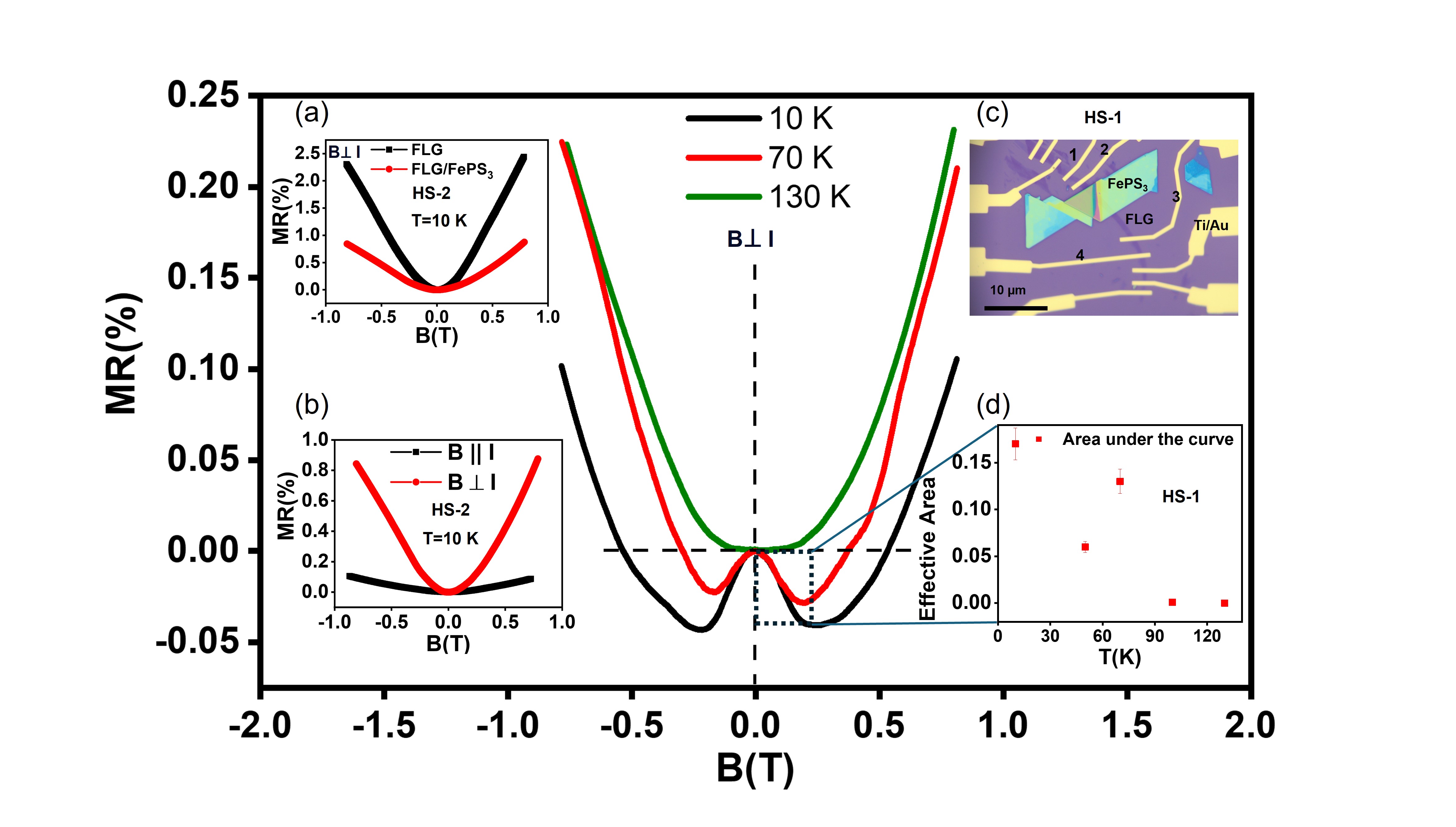}}
\caption{Magnetoresistance (MR, in \%) as a function of magnetic field ($B$) for various temperatures ($T$). Few layer graphene (FLG) exhibits negative MR ($-0.2 \, \text{T} \leq B_\perp \leq 0.2 \, \text{T}$) up to 100 K in HS-1 when measured between electrodes 2 and 3. For temperatures above 100 K, FLG displays positive MR across the entire magnetic field range. Inset (a) presents MR measurements for FLG and the FLG/FePS$_3$ heterostructure (HS-2) in a $B \perp I$ configuration at 10 K, revealing that MR decreases by more than a factor of two in the heterostructure due to the presence of FePS$_3$. Inset (b) shows MR of FLG in HS-2 at the lowest temperature ($T \sim 10 \, \text{K}$) for two different configurations, indicating negligible MR in the $B \parallel I$ configuration. Inset (c) provides an optical microscopy image of the FLG/FePS$_3$ heterostructure (HS-1), with measured electrodes labeled as 1-4; the scale bar indicates the lateral dimensions of the heterostructure. Inset (d) illustrates the area under the curve in the negative MR region, demonstrating a decrease in effective area with increasing temperature up to 100 K, after which it approaches zero.  \label{fig3}} 
\end{figure*}

\begin{figure*}[ht]
\centerline{\includegraphics[scale=0.6, clip]{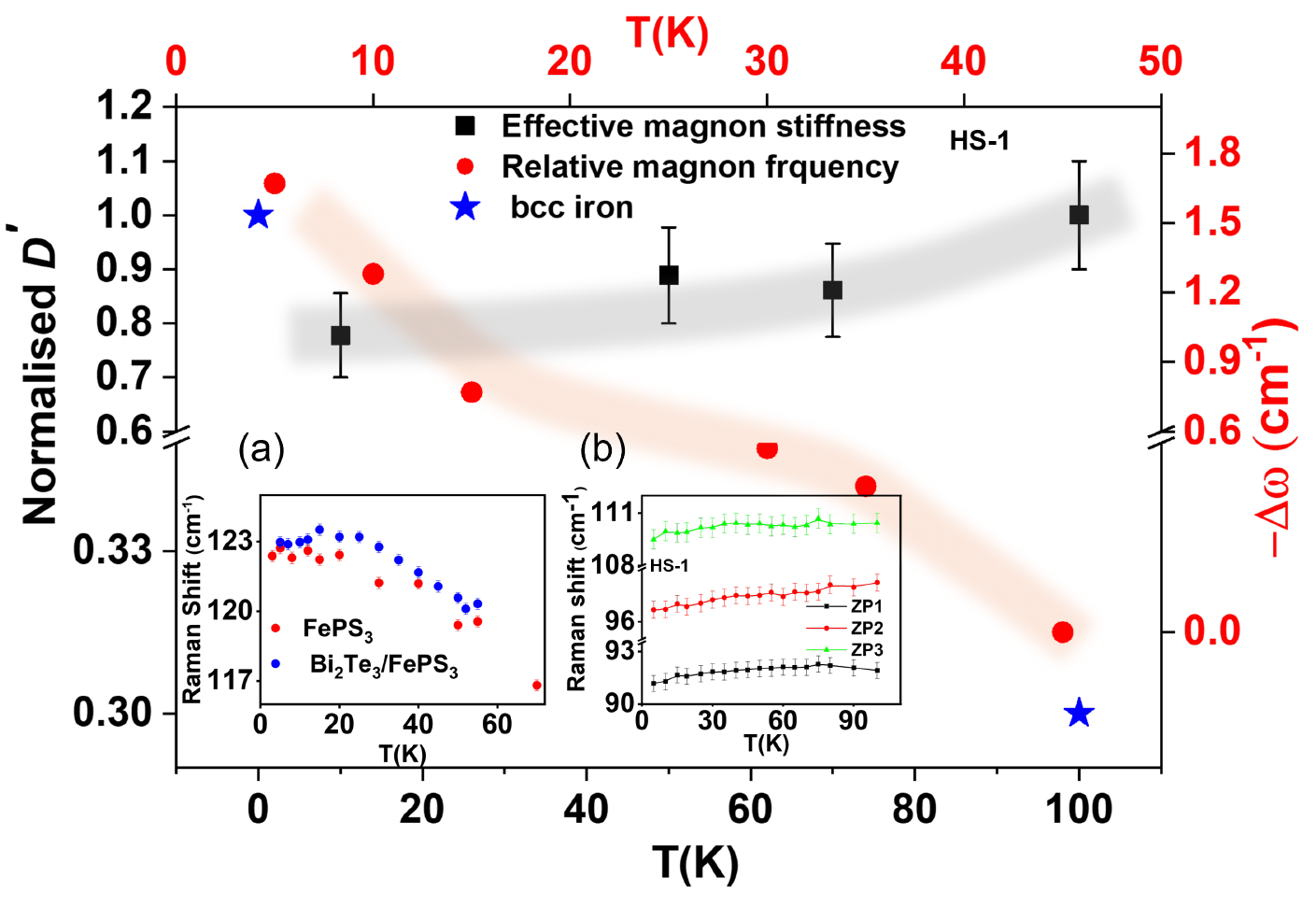}}
\caption{{Normalized magnon stiffness ($D'$) and relative magnon frequency ($\Delta \omega$) as functions of temperature in HS-1. The relative magnon frequency is calculated by subtracting all frequencies from the highest frequency value. In HS-1, the relative magnon frequency increases with temperature, while $D'$ also increases with temperature and remains within the shaded region. For comparison, the magnon stiffness of bcc iron is plotted at two different temperatures alongside our normalized stiffness values. Inset (a) illustrates the anomalous temperature dependence of magnon modes in pristine FePS$_3$ and Bi$_2$Te$_3$/FePS$_3$ heterostructures. Inset (b) presents the temperature-dependent ZP modes. ZP1 modes exhibit a phonon anomaly around 75 cm$^{-1}$ with $\Delta \omega \sim 1$ cm$^{-1}$, whereas ZP2 and ZP3 modes show a fairly linear dependence on temperature with Raman shifts of $\sim 3.27$ cm$^{-1}$ and $\sim 1$ cm$^{-1}$, respectively.} \label{fig4}} 
\end{figure*}

First, we investigate the modification of magnetoelastic and zone-folded phonons in the HS through Raman spectroscopy. FePS$_3$ exhibits three spin-phonon coupled Raman modes (SP1, SP2, SP3) at higher wavenumbers ($\geqslant$ 250 cm$^{-1}$), a magnon mode ($\sim$ 120 cm$^{-1}$) as depicted in Figure \ref{fig2}(a), and three zone-folded peaks at low temperatures  ($T_\text{N}$)(see inset of Figure \ref{fig2}(a)) \cite{PhysRevB.103.064431, PhysRevB.104.134437}. In FLG/FePS$_3$ HS, shifts in the G and 2D bands of FLG indicate charge transfer at the interface ruling out the strain effects (Fig. SI) \cite{yang2019all, lee2012optical,zhang2019study}. Note that  $p$-doping of graphene in the HS can be inferred from the band diagram depicted in Fig. SI \cite{ramos2022photoluminescence,shi2010work}. Phonon anomalies in SP modes appear around 120 K, near the $T_\text{N}$, with stronger deviations in heterostructures ($\Delta\omega$ $\sim$ 1.42 cm$^{-1}$) compared to FePS$_3$ on SiO$_2$ substrates ($\Delta\omega$ $\sim$ 0.44 cm$^{-1}$), suggesting enhanced spin-phonon coupling due to a sharp interface (Fig. \ref{fig2}(b), and see Supplementary Information Fig. SII, Fig. SIII  for details). Moreover, increased phonon linewidth of the G-band with cooling can be correleted to electron-phonon coupling \cite{PhysRevLett.99.176802}. The coefficient of spin-phonon coupling ($\lambda_{sp}$) for the mode around 283 cm$^{-1}$ increases by 4.75 times on sifting from SiO$_2$ to FLG (see Supplementary Information, Table SI), confirming the role of FLG as a substrate for better magnetoelastic coupling \cite{casto2015strong, PhysRevB.100.224427, calder2015enhanced}. Also, contrasting temperature dependent softening of the three zone-folded peaks (ZP1 $\sim $ 90 cm$^{-1}$, ZP2 $\sim $ 97 cm$^{-1}$, and ZP3 $\sim $ 110 cm$^{-1}$) in the HS has been depicted in the inset of Fig. \ref{fig4}(b) \cite{PhysRevB.103.064431}.
Next, we perform magnetotransport measurements of the HS device across the magnetic transition temperature of the AFMI, including temperatures below, at, and above the transition. The temperature dependence (10 K to 150 K) of longitudinal resistivity was measured in the presence of a perpendicular magnetic field ($B_\perp$) ranging from -1 T to 1 T. The magnetoresistance (MR), defined as $[\frac{R (B)-R (0)}{R(0)}]\times 100$, is shown in Fig. \ref{fig3} for HS at different temperatures. In HS-1, FLG exhibits negative MR at low fields (- 0.2 T $\leqslant B_{\perp} \leqslant$ 0.2 T) when measured between electrodes 2 and 3 (see Figure \ref{fig3}), in contrast to the behavior recently reported for graphene/MnPSe$_3$ \cite{PhysRevB.108.125427}. At higher fields ($B> 0.2$ T), MR shows a linear dependence on $B$, consistent with \cite{PhysRevLett.97.016801}. Notably, positive MR was observed in FLG between electrodes not containing FePS$_3$ across the entire field range at temperatures from 10 K to 150 K (see Supplementary Information, Fig. SIV(a)). Furthermore, the negative MR region shrinks with temperature, disappearing above 100 K (see inset Fig. \ref{fig3}). Also, similar MR behavior, with monotonic increase in zero-field resistance, was observed with sweeping gate voltage ($\pm$40 V), linked with the hole-doped nature of graphene. In HS-2, usual positive MR, instead of negative MR, in FLG was observed across the entire magnetic field and temperature range (see Supplementary Information, Fig. SIV(b) and SV). At 10 K, MR increased by ~8\% with decreasing FePS$_3$ thickness, while the overall resistance in FLG increased more than 4 times in the heterostructure (see Supplementary Information, Fig. SIV(c)). As depicted in the inset of Figure \ref{fig3}, the presence of FePS$_3$ in the HS has caused reduction of MR by ~2.77\%, contrasting with the results from graphene/Eu/Si HS thin films \cite{https://doi.org/10.1002/smll.202301295}. Negligible MR was observed with in-plane magnetic fields ($B_\parallel$), likely due to diminishing Lorentz force effects in thin graphene layers (see inset Figure \ref{fig3}(b)). 

The dominant interaction in the present device can be linked to prior studies. MR suppression has been attributed to defects \cite{PhysRevLett.89.266603,PhysRevLett.97.036802}, exfoliation-induced ripples \cite{doi:10.1126/science.1102896}, and mesoscopic corrugations \cite{PhysRevLett.97.016801}. Nitrogen-doped graphene shows negative MR at 2.3 K and 279 K due to defect boundary scattering \cite{doi:10.1021/nn5057063}, while Zhou \textit{et al.} observed a positive-to-negative MR transition in disordered graphene \cite{zhou2011positive}. In large-area FLG, negative MR typically disappears above 30 K \cite{cao2010large}. In this study, minimal defects in FLG are confirmed by the absence of a D-peak in the Raman spectrum. Negative MR persists up to 100 K, ruling out weak localization. Instead, spin-dependent scattering at the interface, consistent with Co nanosheet/graphene interfaces \cite{majumder2020transition}, likely drives the observed behavior.

To understand the negative MR in HS-1, we measured $R$ ($\frac{V_{23}}{I_{12}}$) \textit{vs.} $T$ (see Figure \ref{fig3} for the connections and the Supplementary Information Fig. SVI for the plot) where a slope change at 100 K, near magnon temperature, can be observed. Below 100 K, surface electrons of FLG is likely to get coupled with AFM magnons at the interface and possibly cause negative MR, also observed in magnon modes in FM SrRuO$_3$ \cite{PhysRevLett.123.017202}. But above 100 K, phonon-dominated transport becomes active, which may lead to a positive MR. As the thickness of FePS$_3$ decreases (HS-2), EMC weakens, suppressing negative MR \cite{PhysRevX.9.011026}. As noted earlier, unlike the graphene/MnPSe$_3$ heterostructure, our resistivity data shows no evidence of the Kondo effect, likely due to the low impurity levels confirmed by Raman spectroscopy \cite{PhysRevB.108.125427}.

The resistivity in such systems can be expressed as  $\Delta\rho\propto\frac{BlnB}{D(T)^{2}}$, where $D$ is the magnon stiffness, which decreases with temperature, indicating a less rigid spin system \cite{PhysRevB.66.024433}. We fit our MR data to this formula, and an effective magnon stiffness ($D^{'}$) were plotted with temperature in Fig. \ref{fig4}, showing a consistent decline under cooling. While magnon stiffness can be measured by inelastic neutron scattering (INS), measuring the same in the HS framework is beyond the scope of our study. In pristine FePS$_3$ ($T_\text{N}\sim$ 120 K), magnetization and magnon stiffness both increase with cooling below 60 K \cite{PhysRevB.103.064431, PhysRevB.108.L060403}. However, as illustrated in Fig. \ref{fig4}, in the FLG/FePS$_3$ HS, the magnon mode softens below 70 K (see Supplementary Information, Fig. SVII), suggesting weakened magnetization and possible restoration of the antiferromagnetic ground state \cite{PhysRevB.108.L060403}. Despite these effects, no change in $T_\text{N}$ was observed in FePS$_3$ in the HS with FLG, unlike the Bi$_{2}$Te$_{3}$/FePS$_{3}$ system where magnon stiffness decreased significantly due to the Bi$_{2}$Te$_{3}$ layer \cite{maity2024manipulating}.

A phenomenological model of carrier density fluctuations from electron-magnon coupling, leading to magnon-to-charge conversion, may explain the negative MR below 100 K in HS-1.  Consistent with the strength of the magnetic moment, reducing the FePS$_3$ thickness further suppresses negative MR in HS-2. Theoretical predictions suggest that at the TI/antiferromagnetic insulator interface — comparable to the present heterostructure with its uncompensated interface — charge-magnon interactions can induce magnon-to-charge conversion \cite{PhysRevB.102.115152, PhysRevLett.113.057601}. This can be described by the following Hamiltonian \cite{PhysRevB.104.014508, PhysRevB.105.184434} of electron-magnon coupling expressed by exchange interaction ($J_{em}$), facilitating angular momentum transfer between electron spins and magnons (see the schematic of the Figure \ref{fig1}a).

\begin{equation}
 H_{em} = -J_{em} \sqrt{\frac{S}{2N}}\sum_{\mathbf{k} \mathbf{q}} \Gamma_{\mathbf{q}} c_{\mathbf{k+q}\downarrow}^{\dagger} c_{\mathbf{k} \uparrow} + \Gamma_{\mathbf{q}}^{\dagger} c_{\mathbf{k} + \mathbf{q}\uparrow}^{\dagger} c_{\mathbf{k} \downarrow}    
\end{equation}\label{Hem}
with
\begin{equation*}
\Gamma_{\mathbf{q}} = \left( \gamma_A u_{\mathbf{q}} + \gamma_B v_{\mathbf{q}} \right) a_{\mathbf{q}} + \left( \gamma_A v_{\mathbf{q}} + \gamma_B u_{\mathbf{q}} \right) b_{\mathbf{-q}}^{\dagger}    
\end{equation*}

where, $c_{\mathbf{k}}^{\dagger}$, $c_{\mathbf{k}}$ are the creation and annihilation operators for the electron, and $a_{\mathbf{q}}^{\dagger}$, $a_{\mathbf{q}}$, $b_{\mathbf{q}}^{\dagger}$, $b_{\mathbf{q}}$ denote the same for magnon in two sublattices $A$ and $B$. $S$ is the spin per atom of the magnetic layer, and $N$ is the number of atomic sites at the interface. $J_{em}\gamma_{A}$ and $J_{edm}\gamma_{B}$ are the interfacial exchange coupling with $A$ and $B$ sublattices. The $u_{\mathbf{q}}$ and $v_{\mathbf{q}}$ are two prefactors used in Bogoliubov transformation which satisfies $u_{\mathbf{q}}^2-v_{\mathbf{q}}^2=1$.\\

Under non-equilibrium conditions, applying an electric field or magnetic field may lead to magnon accumulation/depletion ($\delta n_m$), which contributes to the interconnected conductivity ($\sigma_{em}$), characterized by the non-equilibrium electron and magnon distribution functions $f_\sigma(\mathbf{x,k})$, and  $g^m(\mathbf{x,q})$, .
Therefore, with the help of linearised Boltzmann equation in the description of electronic transport with group velocity $v_{k_x}$, the additional current induced by the exchange coupling could be expressed as $j_{em} \sim \int d\mathbf{k} v_{k_x} \sum_{\sigma}\tau_\sigma\left[ \frac{\partial f_\sigma(x,\mathbf{k})}{\partial t} \right]_{em}$ , which  is directly proportional to the $J_{em}^2$ \cite{PhysRevB.105.184434}, where $\tau_\sigma$ is the electronic relaxation time for  spin $\sigma$ (see supplementary Information for details).  
Note that $J_{em}^2$ dependence remains consistent when a thermal gradient is applied across the magnetic insulator \cite{PhysRevB.86.214424}.

We now address the magnetic field dependence of magnon dispersion and connection to negative MR. When an external magnetic field ($H$) is applied, it alters the energy landscape of magnons by opening a magnon gap, modifying magnon energy as $\epsilon_{\mathbf{q}} \rightarrow \epsilon_{\mathbf{q}}' = \epsilon_{\mathbf{q}} \pm g\mu_{\text{B}}H $, resulting in more defined long-wavelength magnons and increased magnon stiffness. This revision of the magnon chemical potential alters the distribution function, influencing magnon accumulation ($\delta n_m$). Consequently, a reduced spin-wave damping enhances magnon coherence \cite{PhysRevLett.122.187701}, leading to well-defined momentum and phase, which improves momentum transfer to conduction electrons, thus strengthening the EMC and self-energy effects, as evidenced from our Raman spectroscopy data \cite{PhysRevB.104.125125}. The momentum transfer facilitates electron movement along the applied electric field, enhancing net current and conductivity. Moreover, the magnetic field can also increase magnetic anisotropy, reinforcing the exchange interactions at the interface, which promotes efficient magnon-to-charge conversion \cite{PhysRevResearch.5.L022065}. At low temperatures, reduced thermal excitation of magnons minimizes random scattering, maintaining coherence.

 Importantly, interfacial charge transfer enhances the orbital overlap, and the exchange coupling between Fe$^{2+}$ ions in FePS$_3$ \cite{PhysRevResearch.5.L022065}. Accordingly, greater lattice softening occurs below the magnetic ordering temperature ($T_\text{N}$), increasing the HS's susceptibility to dielectric property changes. 

\begin{figure*}
%\centerline{\includegraphics[width=\linewidth]{F5.jpg}}
\centerline{\includegraphics[scale=0.6, clip]{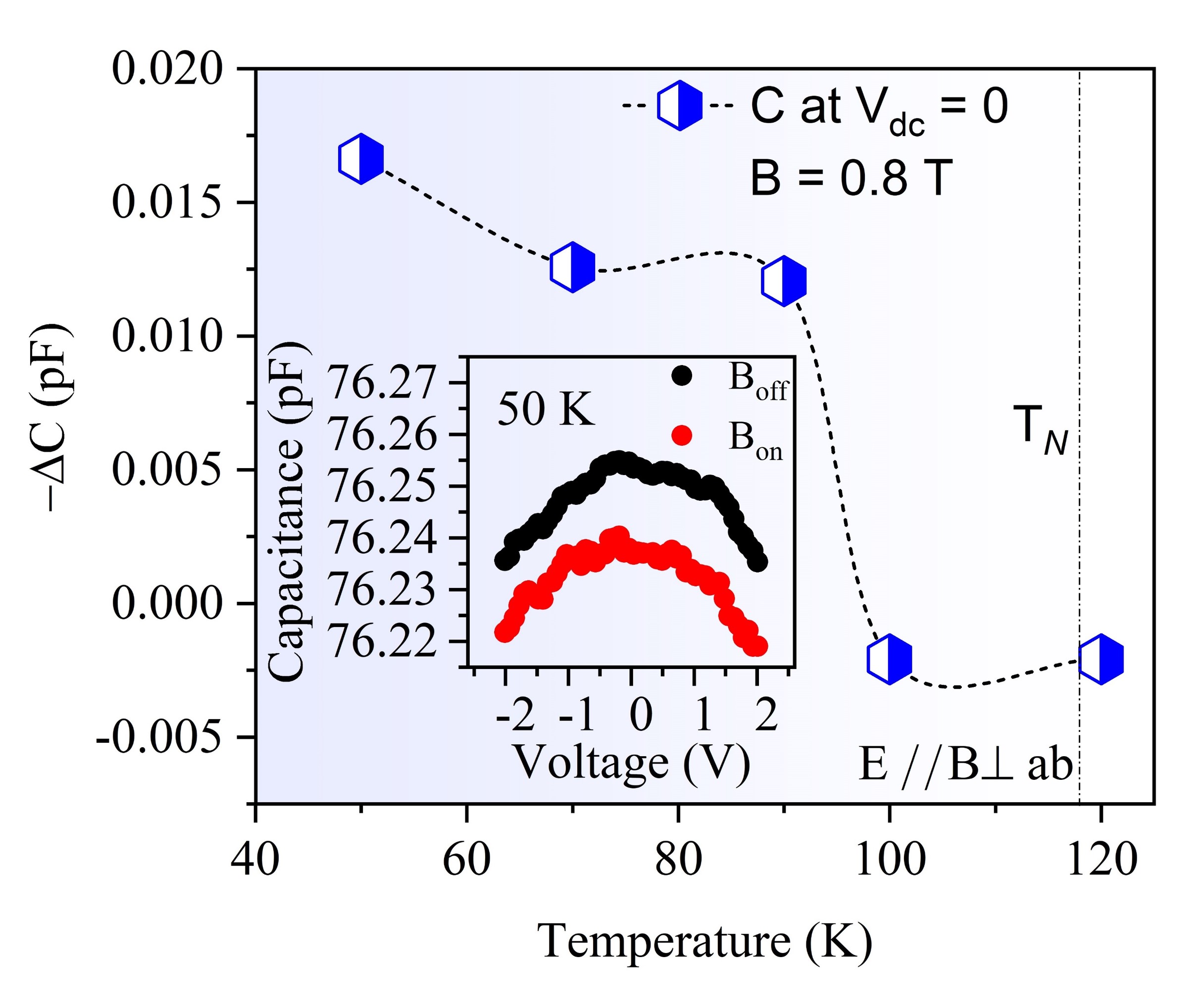}}
\caption{\label{fig:epsart} Temperature dependence of change in capacitance with applied magnetic field below T$_N$. The change in capacitance increases at significantly lower temperatures than T$_N$, from which magnon modes become prominent. Inset shows capacitance with applied dc voltage on FLG/FePS$_3$ /FLG device measured at 100 kHz frequency at 50 K temperature with and without applying magnetic fields perpendicular to the \textit{ab}-plane. Upon the application of magnetic fields, the overall capacitance has been decreased. In the presence of magnetic fields, no significant changes have been observed in the trend of capacitance with the applied dc bias.    
\label{CV}}
\end{figure*}

To investigate the interfacial effects in detail, further C-V measurements were conducted on the FLG/FePS$_3$/FLG HS with electric field, and magnetic field applied perpendicular to the \textit{ab}-plane of the FePS$_3$ flake at selected temperatures above and below $T_\text{N}$. The C-V characteristics displayed typical diminutive behavior with voltage, maintained under magnetic fields  \cite{brennan1992characterization, boni2015electrode}. Notably, a significant drop in capacitance (\(-\Delta C \sim 15 \, \text{fF}\) at 20 K) was observed with the application of the magnetic field ($\pm$ 1 T), indicative of negative magnetocapacitance (or magnetodielectric effect), commonly seen in zigzag antiferromagnetic insulators \cite{PhysRevB.105.054408, huang2022antiferromagnetic, PhysRevB.108.L060403} (see Figure \ref{CV}). While MD coupling is usually detected near magnetic ordering temperatures \cite{lawes2009magnetodielectric}, the observed capacitance drop shows an upturn at 90 K, well below the $T_\text{N}$ (\(\sim 120 \, \text{K}\)), implying an additional mechanism at play. The appearance of the magnon mode at 90 K suggests a possible correlation with magnon-phonon hybridization (indicated by magneto-Raman data) and the negative magnetocapacitance effect \cite{PhysRevB.108.085435,PhysRevB.104.134437}. Enhancement of magnon-to-charge conversion under magnetic field may influence the dielectric displacement, allowing capacitance to be further adjusted, as schematically shown in Fig. \ref{fig1}b. In the future, the interaction between elementary excitations, such as spin waves and electric polarization, could facilitate the development of magnon capacitors, as theoretically proposed in several studies \cite{semenov2010graphene,datta2005proposal,gunnink2024magnon}.

\section{Conclusion} 

In conclusion, the combination of low-temperature Raman scattering, magnetoresistance, and magnetocapacitance measurements suggests a potential magnon-to-charge conversion at the interface of FLG/ antiferromagnetic FePS\(_3\) heterostructure. The three spin-phonon coupled modes of FePS\(_3\) exhibit a departure from anharmonic behavior around 120 K, similar to its pristine form; however, the deviation from anharmonicity, \(\Delta\omega\), increases threefold in the presence of FLG. The linewidths of these modes follow a three-phonon anharmonic decay process. In the heterostructure, the magnon mode of FePS\(_3\) experiences a blueshift in frequency with increasing temperature, while magnon stiffness decreases as the temperature drops. Magneto-transport measurements reveal that FLG exhibits negative magnetoresistance (MR) at low fields (\(\pm 0.2 \, \text{T}\)), transitioning to positive MR for \(B > 0.2 \, \text{T}\). The origin of the negative MR is not attributed to defects, as noted in the literature, but rather to interfacial effects arising from surface electrons in FLG coupling with magnons in FePS\(_3\), sustaining negative MR around 100 K. This magnon-to-charge conversion enhances the electron-magnon coupling strength at the interface below 100 K. Above this temperature, FLG shows positive MR due to phonon-dominated transport. Notably, the magnon-to-charge conversion diminishes with decreasing layer thickness, leading to reduced electron-magnon coupling and suppression of negative MR in heterostructures with thinner FePS\(_3\). Moreover, the negative magnetocapacitance effect in the FLG/FePS\(_3\)/FLG heterostructure indicates a coupling between magnons and interfacial electric displacement. Our results may pave the way for electric field control of magnons, enabling the development of magnon-based logic devices.

\section*{Acknowledgements}
The authors thank Dr. Vasant Sathe, Prof. Biswajit Karmakar, and Dr. Mintu Mondal for their generous cooperation during the experiments. S. M. and T. K. are grateful to DST-INSPIRE for the fellowship and S. M. acknowledges Dr. Pooja Agarwal and Suvankar Purkait for their experimental assistance. S. D. acknowledges Prof. J. Suffczy{\'n}ski for experimental help and appreciates Dr. K. D. M. Rao's efforts as the faculty in charge of e-beam lithography at IACS. K. D. acknowledges TRC for the fellowship. M. P. and B. D. thank IACS for their fellowships. S. D. acknowledges financial support from DST-SERB grant No. CRG/2021/004334, as well as support from the Central Scientific Service (CSS) and the Technical Research Centre (TRC), IACS, Kolkata.

\bibliography{cite}

\end{document}

\title{Supplementary Information: Electron-Magnon Coupling Mediated Magnetotransport in Antiferromagnetic van der Waals Heterostructure}

\author{Sujan Maity}
\affiliation{School of Physical Sciences, Indian Association for the Cultivation of Science, 2A \&  2B Raja S. C. Mullick Road, Jadavpur, Kolkata - 700032, India}
%\email{}
%\affiliation{School of Physical Sciences, Indian Association for the Cultivation of Science, 2A \& B Raja S. C. Mullick Road, Jadavpur, Kolkata - 700032, India}

\author{Soumik Das}

\affiliation{School of Physical Sciences, Indian Association for the Cultivation of Science, 2A \& 2B Raja S. C. Mullick Road, Jadavpur, Kolkata - 700032, India}

\author{Mainak Palit}

\affiliation{School of Physical Sciences, Indian Association for the Cultivation of Science, 2A \& 2B Raja S. C. Mullick Road, Jadavpur, Kolkata - 700032, India}

\author{Koushik Dey}
\affiliation{School of Physical Sciences, Indian Association for the Cultivation of Science, 2A \& 2B Raja S. C. Mullick Road, Jadavpur, Kolkata - 700032, India}

\author{Bikash Das}
\affiliation{School of Physical Sciences, Indian Association for the Cultivation of Science, 2A \& 2B Raja S. C. Mullick Road, Jadavpur, Kolkata - 700032, India}

\author{Tanima Kundu}
\affiliation{School of Physical Sciences, Indian Association for the Cultivation of Science, 2A \& 2B Raja S. C. Mullick Road, Jadavpur, Kolkata - 700032, India}

\author{Rahul Paramanik}
\affiliation{School of Physical Sciences, Indian Association for the Cultivation of Science, 2A \& 2B Raja S. C. Mullick Road, Jadavpur, Kolkata - 700032, India}

%\author{Suvankar Purkait}
%%\affiliation{Experimental Condensed Matter Physics Division,
%%Saha Institute of Nuclear Physics, HBNI, 1/AF Bidhannagar, Kolkata 700064, India}
%\affiliation{Saha Institute of Nuclear Physics, HBNI, 1/AF Bidhannagar, Kolkata 700064, India}

\author{Binoy Krishna De}
\affiliation{UGC-DAE Consortium for Scientific Research, Indore Centre, University Campus, Khandwa Road, Indore, 452001 India}
\author{ Hemant Singh Kunwar}
\affiliation{UGC-DAE Consortium for Scientific Research, Indore Centre, University Campus, Khandwa Road, Indore, 452001 India}

%\author{J. Suffczy{\'n}ski}
%\affiliation{Faculty of physics, University of Warsaw, Pasteura 5, Warsaw, Poland}

\author{Subhadeep Datta*}
\affiliation{School of Physical Sciences, Indian Association for the Cultivation of Science, 2A \& 2B Raja S. C. Mullick Road, Jadavpur, Kolkata - 700032, India}
\email{sspsdd@iacs.res.in}
\maketitle

\begin{figure}[H]
\centerline{\includegraphics[scale=0.5, clip]{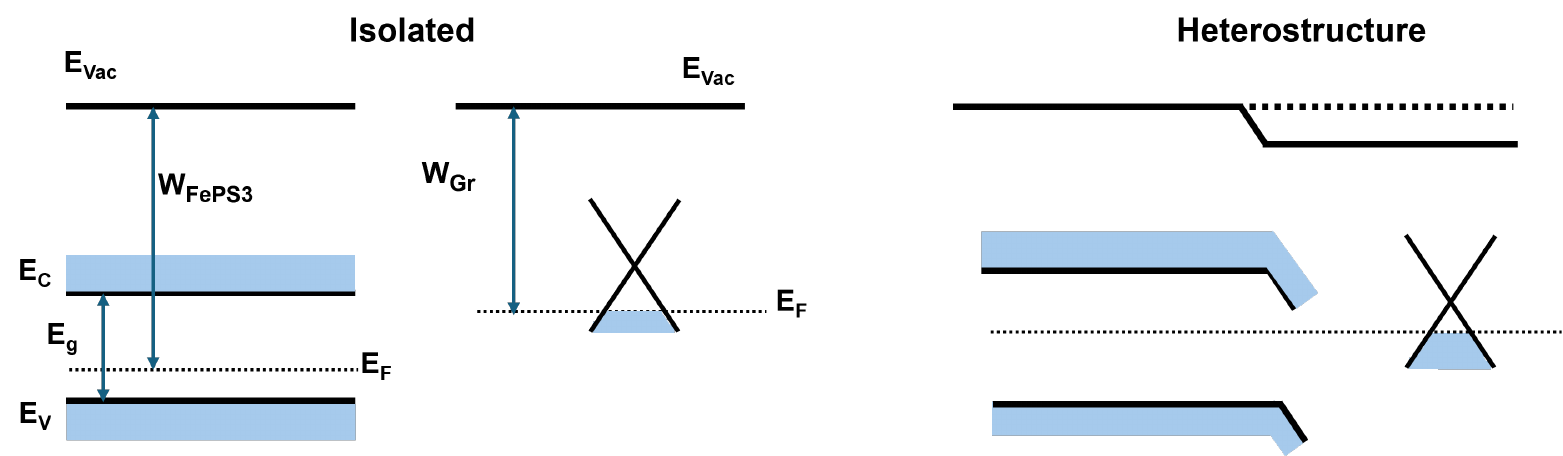}}
\caption{A schematic representation of the energy levels for isolated FePS$_{3}$ and graphene before and after contact. The work functions ($W$) for FePS$_{3}$ and graphene are 4.9 eV and 4.66 eV respectively. E$_{C}$, E$_{V}$, E$_{F}$ and E$_{g}$ are denoted as conduction band, valence band, Fermi level and bandgap respectively. Band bending occurs to equilibrate the Fermi level of the two materials resultting electron transfer from graphene to FePS$_{3}$. Graphene becomes heavily p-doped.}
\end{figure}
\begin{figure}[H]
\centerline{\includegraphics[scale=0.5, clip]{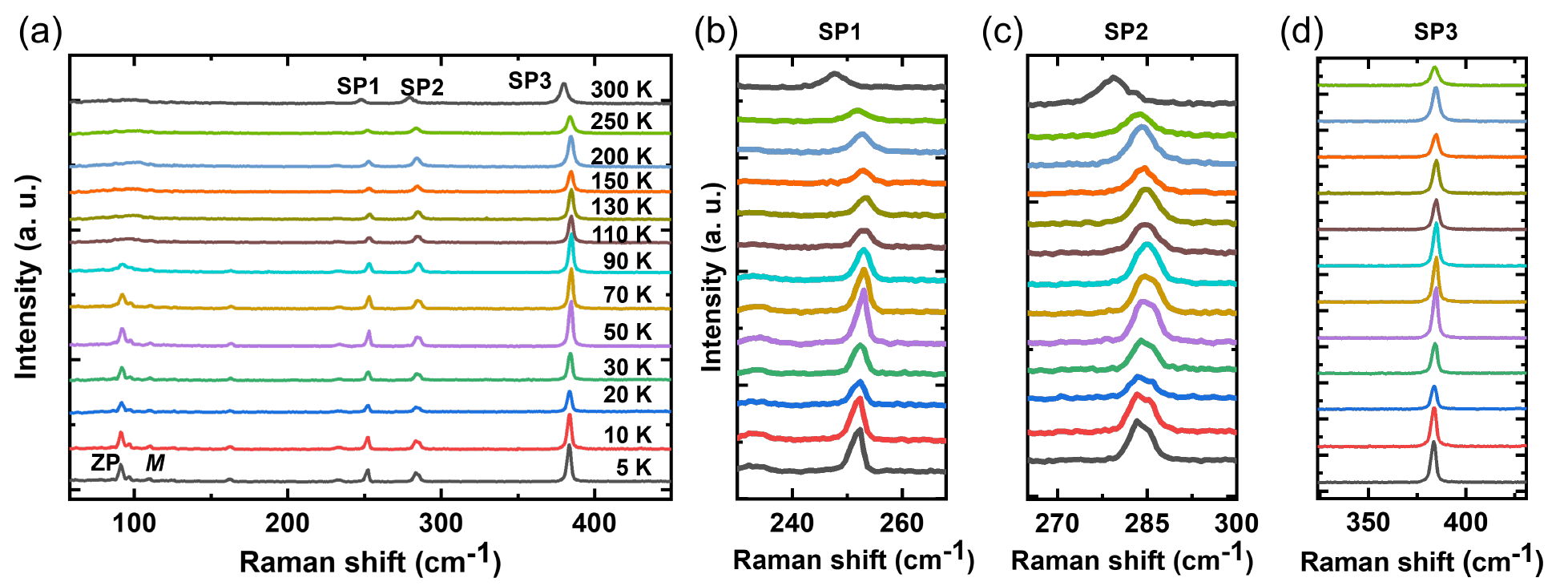}}
\caption{(a) Evolution of different Raman peaks of FePS$_{3}$ with temperature varying from 5 K to 300 K; (b)-(d) Three SP peaks are plotted separately with varying temperature (5 K - 300 K).} 
\end{figure}
\begin{figure}[H]
\centerline{\includegraphics[scale=0.5, clip]{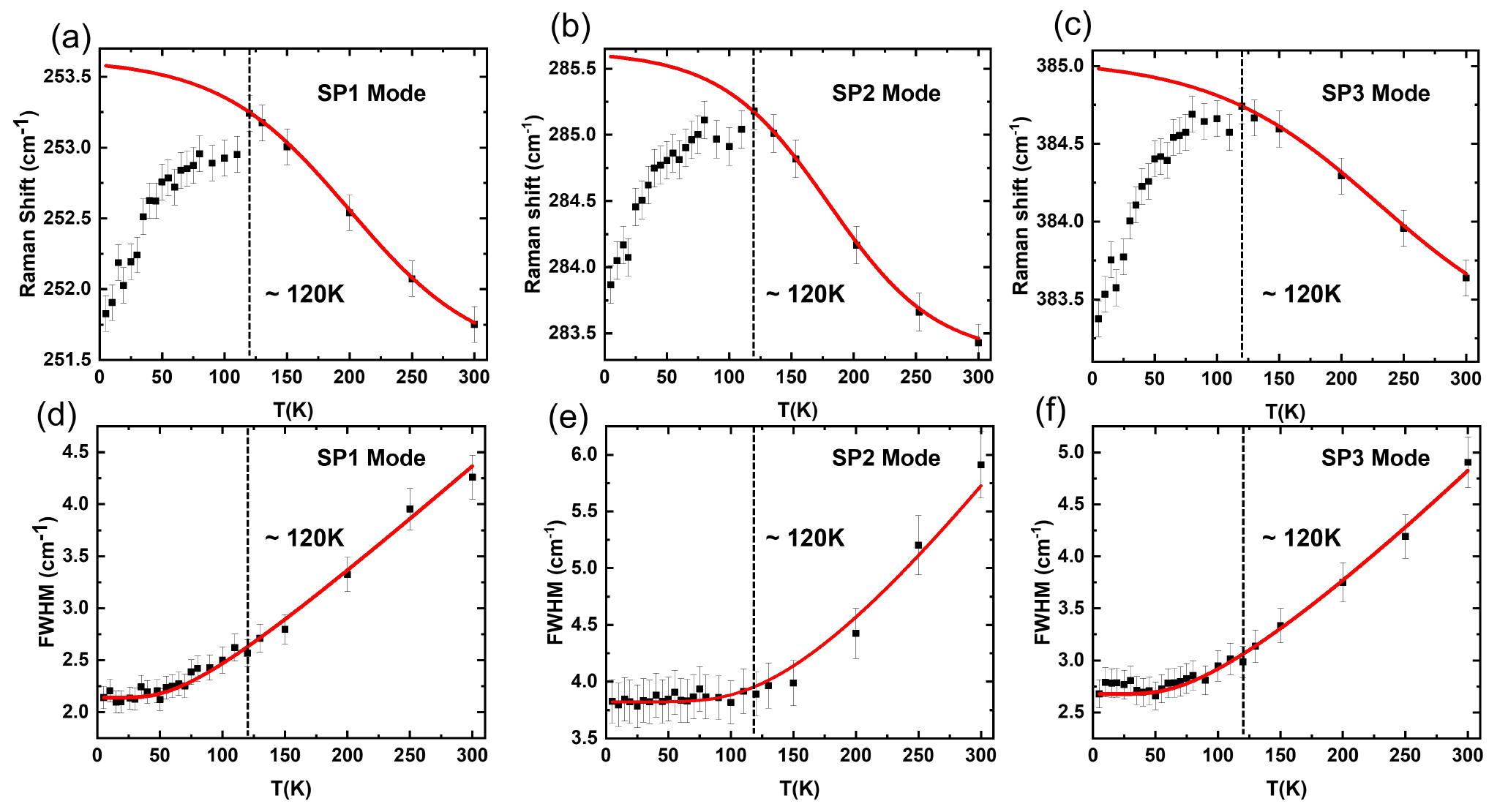}}
\caption{Variation of Peak position and linewidth of FePS$_{3}$ Raman modes (a), (d) SP1 mode; (b), (e) SP2 mode; (c),(f) SP3  mode in FLG/FePS$_{3}$ heterostructure (HS-1) with temperature. All spin-phonon peaks show phonon anomaly at/around 120 K in phonon frequency. Linewidths of all spin-phonon coupled modes obey three-phonon anharmonic decay with temperature.} 
\end{figure}
\begin{figure}[H]
\centerline{\includegraphics[scale=0.5, clip]{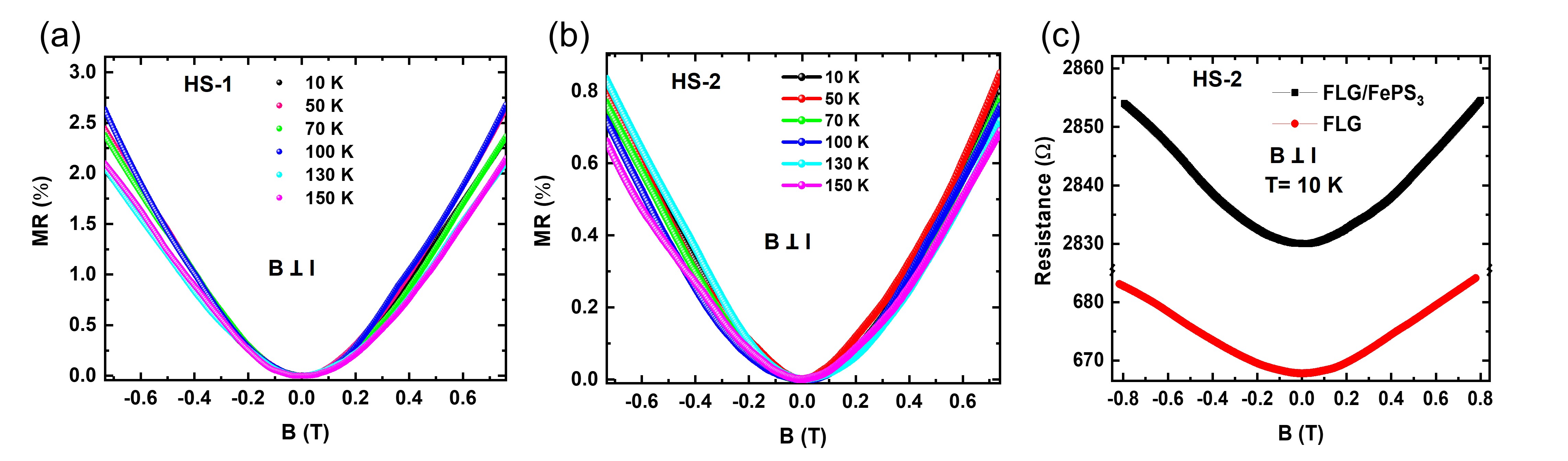}}
\caption{(a) FLG shows positive MR in HS-1 at whole temperature window when mesasured between 1, 2 or 3, 4. (b) Suppression of negative MR is observed in FLG when supported to FePS$_{3}$ with reduced thickness in HS-2 , when measured between 2, 3. (c) Difference in resistance indicates the formation of interface. Overall resiatance of FLG increases $\sim$ 4 times due to FePS$_{3}$ at 10 K.}
\end{figure}
\begin{figure}[H]
\centerline{\includegraphics[scale=0.5, clip]{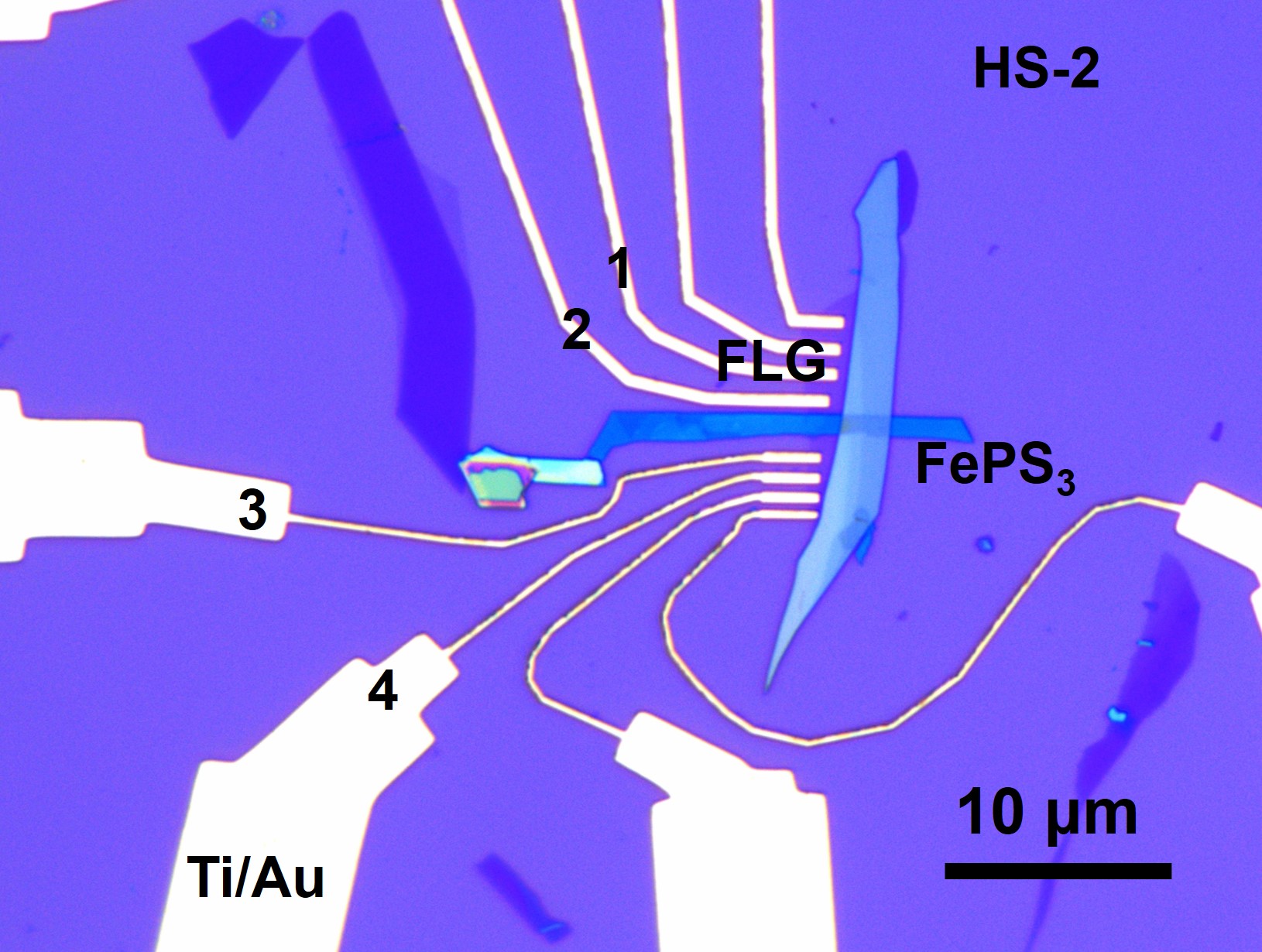}}
\caption{Optical microscopy image of FLG/FePS$_{3}$ heterostructure with reduced thickness (HS-2). Suppresion of negative MR was observed with reduced thickness of FePS$_{3}$ when measured between 2, 3 attributed to the reduction of coupling strength at the interface of the heterostructure.}
\end{figure}
\begin{figure}[H]
\centerline{\includegraphics[scale=0.5, clip]{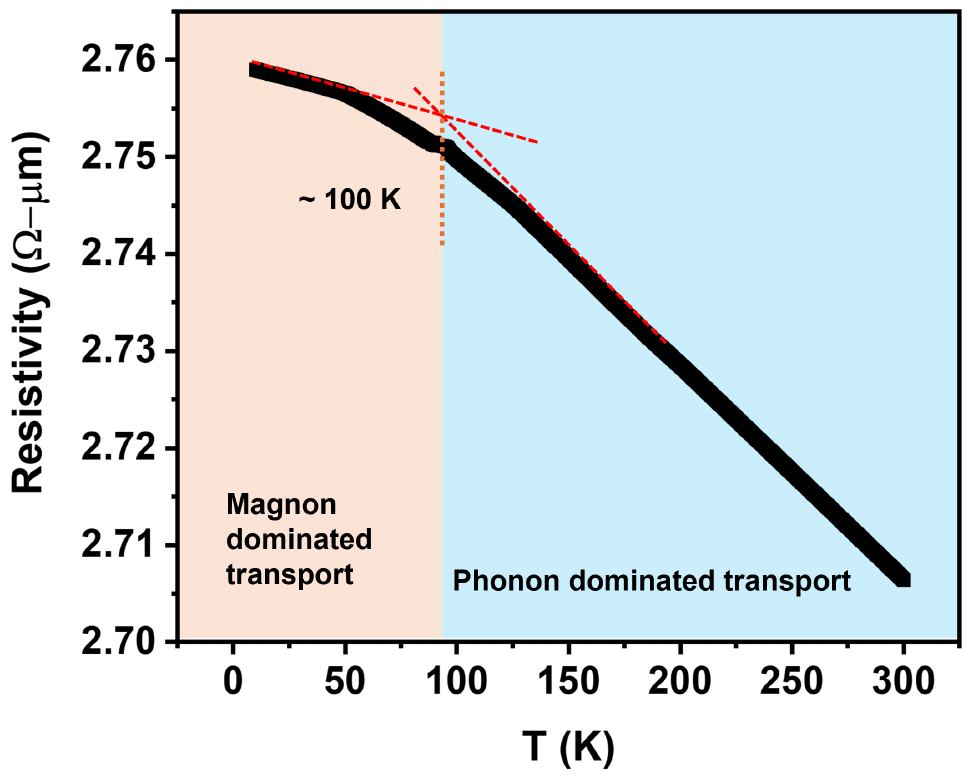}}
\caption{Four terminal longitudinal resistivity as a function of temperature in HS-1. 10 $\mu$A current was applied between electrode 1 and 4. Voltage was measured between 2, 3. The slope changes at/around 100 K and it is attributed to the coupling between surface electrons of FLG and the magnon of FePS$_{3}$. Above 100 K, phonon dominated transport was observed.}
\end{figure}
\begin{figure}[H]
\centerline{\includegraphics[scale=0.5, clip]{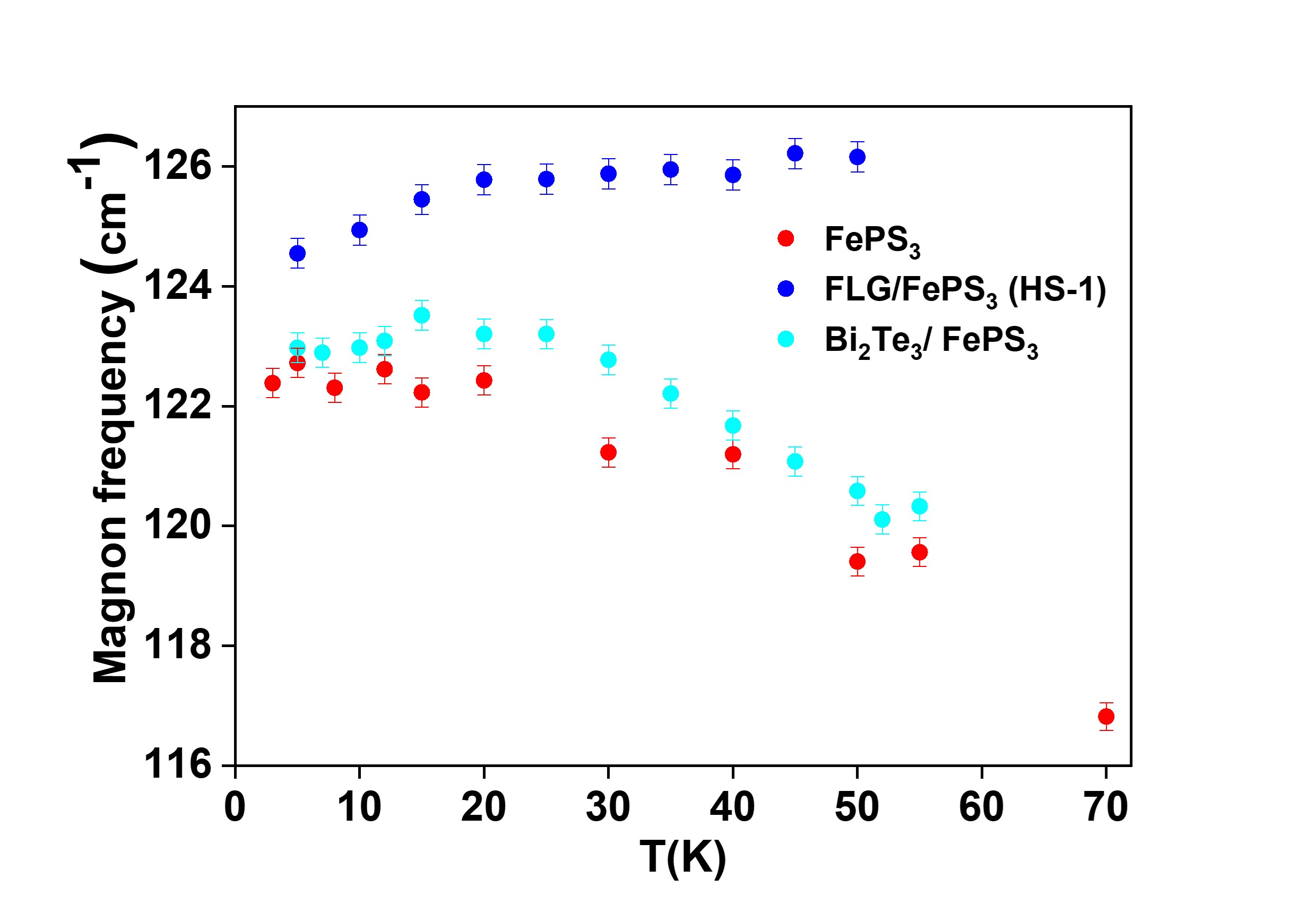}}
\caption{Magnon frequency of FePS$_{3}$ and different heterostructures with varying temperatures. Magnon mode gets blueshifted in frequency with temperature in HS-1. In case of pristine FePS$_{3}$ and Bi$_{2}$Te$_{3}$/FePS$_{3}$ heterostructure, Magnon mode show anomalous behaviour in frequency unlike HS-1.}
\end{figure}
\centering
\begin{table}[h]
\begin{tabular}{|c|c|c|c|}
\cline{1-3}
%\multicolumn{3}{c}{Bi$_{2}$Te$_{3}$}\\
%\hline
\cline{1-3}
Raman Mode (cm$^{-1})$ &  $\lambda_{sp}$(Pristine FePS$_{3}$) & $\lambda_{sp}$ (HS-1) \\ \cline{1-3}
383 & 0.16 & 0.69 \\ \cline{1-3}
283 & 0.12 & 0.57 \\ \cline{1-3}
251 & 0.15 & 0.50 \\ \cline{1-3}
\end{tabular}
\centering \caption{Calculated values of spin-phonon coupling constant ($\lambda_{sp}$) of pristine FePS$_{3}$ and FLG/FePS$_{3}$ heterostructure (HS-1). $\lambda_{sp}$ value increases  more than 4 times for the mode around 383 and 283 cm$^{-1}$ when the substrate is changed from SiO$_{2}$ to graphene. }
\end{table}

\justify{

\section{Crystal structure} 

%\justify{
FePS$_{3}$ crystallizes into monoclinic symmetry with \textit{C}2/\textit{m} space group \cite{PhysRevX.11.011024} with a large out-of-plane magnetic anisotropy observed by magnetic susceptibility measurement \cite{lee2023giant, PhysRevB.103.064431}.

\section{Raman Spectroscopy of the heterostructure}
Note that FePS$_{3}$ shows layer thickness independent transition temperature from bulk to monolayer with $T_\text{N} \sim $ 120 K \cite{lee2016ising, gibertini2019magnetic}. The prominent G peak at around 1582 cm$^{-1}$ suggests the presence of a hexagonal lattice formed by carbon atoms. The absence of the disorder-induced D peak around 1360 cm$^{-1}$ indicates an exceptionally low density of defects throughout the sample. It is noteworthy that no D peak was detected in any region of the sample. 
%The 2D peak at approximately 2700 cm$^{-1}$ exhibits a lower intensity compared to the G peak and shows a broad linewidth of around 86 cm$^{-1}$ (see Figure \ref{fig1}(c)). 
% 
To better understand the charge transfer at the FLG/FePS$_{3}$ interface, band allignment is analyzed considering the work functions of both materials (4.9 eV for FePS$_{3}$ \cite{ramos2022photoluminescence} and 4.66 eV for graphene  \cite{shi2010work}). A schematic representation of the energy levels for FePS$_{3}$ and graphene before and after contact is shown in Supplementary Information, Figure SI. In the heterostructure, the conduction and valence band edges of FePS$_{3}$ bend downward. 

In our case, temperature dependent phonon linewidth of FePS$_{3}$ supported on FLG is completely dominated by phonon-phonon coupling over electron-phonon coupling. Linewidth of SP (1-3) modes obey three-phonon anharmonic decay model in the whole temperature window (see Supplementary Information SIII (d)-(f)). Theoretically, SP coupled Hamiltonian can be expressed as $H_{sp-ph} = \sum_{\alpha q} \frac{\partial H_{spin}}{\partial Q_{\alpha q}} Q_{\alpha q}$,  where $q$ is the set of reciprocal lattice vectors and $Q_{\alpha q}$ is the ionic displacement for a phonon with a frequency $\omega_{\alpha q}$ \cite{casto2015strong, PhysRevB.100.224427}. In the weak SP coupling approximation, we can write $\Delta \omega = \lambda_{sp} \langle \mathbf{S}_i \cdot \mathbf{S}_{i+1} \rangle = -\lambda_{sp} S^2 \phi(T)$, where $\lambda_{sp}$ is the coefficient of SP coupling, $\phi(T)$ is the order parameter given by $ \phi(T) = 1 - \left(\frac{T}{T_\text{N}}\right)^\gamma$, and $S$=2. The normalized deviation ($\Delta\omega$) as a function of reduced temperature ($\frac{T}{T_\text{N}}$) can be well described by the parameter $\gamma$. Using the above equation, the calculated values of the parameter $\lambda_{sp}$ for the three different SP coupled modes of pristine FePS$_{3}$ \cite{PhysRevB.103.064431} and FLG/FePS$_{3}$ heterostructure are listed in Supplementary Information, Table SI. 
%}

%\section{Magnetocapacitance measurement}

%The capacitance vs. voltage (C-V) measurements were carried out on FLG/FePS$_{3}$/FLG device at some fixed temperatures using a Zurich MFIA impedance analyzer and a CRYOCON (22c) temperature controller. Capacitance have been measured on this heterostructure by superimposing \text{500} mV peak-to-peak ac signal (frequency = \text{100} kHz) with a dc voltage varying up to $\pm$ \text{2} V. At each temperature, magnetic field have been varied up to $\pm$ \text{0.8} T via an electromagnet along the applied voltage direction using a close cycle refrigerator (CCR) and gold contacts have been fixed to the extended FLG from both the surfaces of FePS$_{3}$.

\section{Electron Movement mediated by Magnon at the metal/magnetic insulator interface}

%\justify{
The Hamiltonian of electron-magnon coupling can be expressed by the exchange interaction $J_{em}$, which represents overlapping wave functions of the conduction electrons and the magnetic ions. This exchange interaction facilitates angular momentum transfer between the electron spins on the metallic side and the magnon on the antiferromagnetic insulator side.The electron-magnon coupling can be expressed by the Hamiltonian as follows \cite{PhysRevB.104.014508, PhysRevB.105.184434}:

\begin{equation}
 H_{em} = -J_{em} \sqrt{\frac{S}{2N}}\sum_{\mathbf{k} \mathbf{q}} \Gamma_{\mathbf{q}} c_{\mathbf{k+q}\downarrow}^{\dagger} c_{\mathbf{k} \uparrow} + \Gamma_{\mathbf{q}}^{\dagger} c_{\mathbf{k} + \mathbf{q}\uparrow}^{\dagger} c_{\mathbf{k} \downarrow}    
\end{equation}\label{Hem}
with
\begin{equation}
\Gamma_{\mathbf{q}} = \left( \gamma_A u_{\mathbf{q}} + \gamma_B v_{\mathbf{q}} \right) a_{\mathbf{q}} + \left( \gamma_A v_{\mathbf{q}} + \gamma_B u_{\mathbf{q}} \right) b_{\mathbf{-q}}^{\dagger}    
\end{equation}

Here, $c_{\mathbf{k}}^{\dagger}$, $c_{\mathbf{k}}$ are the creation and annihilation operators for the electron, and $a_{\mathbf{q}}^{\dagger}$, $a_{\mathbf{q}}$, $b_{\mathbf{q}}^{\dagger}$, $b_{\mathbf{q}}$ denote the same for magnon in two sublattices $A$ and $B$. $S$ is the spin per atom of the magnetic layer, and $N$ is the number of atomic sites at the interface. $J_{em}\gamma_{A}$ and $J_{edm}\gamma_{B}$ are the interfacial exchange coupling with $A$ and $B$ sublattices. The $u_{\mathbf{q}}$ and $v_{\mathbf{q}}$ are two prefactors used in Bogoliubov transformation which satisfies $u_{\mathbf{q}}^2-v_{\mathbf{q}}^2=1$.\\

At the interface, due to the presence of electron-magnon coupling, magnon to charge conversion would be possible under non-equilibrium conditions, either by applying electric field in the metal or by applying magnetic field or temperature gradient in AFM insulator.
This would result into a non-local conductance in the metallic layer \cite{goennenwein2015non, li2016observation}, characterized by a non-equilibrium distribution function \cite{ashcroft1976solid}.\\

The distribution functions for electron, $f_\sigma(\mathbf{x,k})$, and magnon, $g^m(\mathbf{x,q})$, in terms of equilibrium and non-equilibrium function could be written as  \cite{PhysRevLett.109.096603, PhysRevB.105.184434, PhysRevB.96.024449, PhysRevB.48.7099, PhysRevB.89.014416}
\begin{equation}
f_\sigma(x, \mathbf{k}) = f_\sigma^0(\mathbf{k}) - \frac{\partial f_\sigma^0(\mathbf{k})}{\partial \epsilon_\mathbf{k}} \left[\delta \mu_\sigma(x) + \eta_\sigma(x, \mathbf{k}) \right]
\end{equation}
\begin{equation}
g^m(x, \mathbf{q}) = g^0(\mathbf{q}) - \frac{\partial g^0(\mathbf{q})}{\partial \epsilon_\mathbf{q}^m} \left[\delta \mu^m(x) + \eta^m(x, \mathbf{q}) \right]
\end{equation}
Here, $f_\sigma^0(\mathbf{k}) = \frac{1}{exp \frac{\epsilon_{\mathbf{k}}-\mu_{\sigma}}{k_{\text{B}} T}+1}$ and $g^0(\mathbf{k}) = \frac{1}{exp \frac{\epsilon^m_{\mathbf{q}}-\mu^m}{k_{\text{B}} T}-1}$ are the equilibrium distribution functions for electron and magnon respectively with electronic enrgy, $\epsilon_{\mathbf{k}}$, and magnon energy, $\epsilon_{\mathbf{q}}^m = Dq^2 +\Delta$. $D$ is the magnon stiffness and $\Delta$ is the magnon gap energy. The functions  $\eta_\sigma(x, \mathbf{k})$ and $\eta^m(x, \mathbf{q})$ capture any variations related to momentum-dependent corrections to the excitation energies, whereas $\delta \mu(x)$ is an uniform shift of the chemical potential.\\

%
%Thus, under non-equilibrium condition, an electron accumulation, 
%$\delta n_s(x) = \frac{1}{\sqrt{2\pi}} \int d\mathbf{k} \left[ f_\uparrow(x,\mathbf{k}) - f_\downarrow(x,\mathbf{k}) \right]$ would introduce a spin-current in the metallic side as $j_s(x) \equiv \frac{1}{\sqrt{2\pi}} \int d\mathbf{k} v_{k_x} \left[ f_\uparrow(x,\mathbf{k}) - f_\downarrow(x,\mathbf{k}) \right]$. Similarly, a magnon accumulation, $\delta n_m(x) = \frac{1}{\sqrt{2\pi}} \int d\mathbf{q} \left[ g(x,\mathbf{q}) - g^0(\mathbf{q}) \right]$ mediate a magnon-current in the magnetic insulator side as
%$j_m(x) \equiv \frac{1}{\sqrt{2\pi}} \int d\mathbf{q} v_{q_x} g(x,\mathbf{q})$.  
%Our focus, however, is on the additional contribution to the electric current density, $j_e$, caused by magnon accumulation, which results in an interconnected conductivity that arises from the interaction Hamiltonian stated above in eq.\ref{Hem}.\\

Thus, under non-equilibrium condition, an electron accumulation would introduce a spin-current in the metallic side as identified by the spin-polarized distribution function. Consequently, a magnon accumulation characterized by the non-equilibrium distribution function over it's equilibrium part mediate a magnon-current in the magnetic insulator side \cite{PhysRevB.96.024449}.   Our focus, however, is on the additional contribution to the electric current density ($j_e \sim \int d\mathbf{k} v_{k_x} \sum_{\sigma} f_\sigma(x,\mathbf{k})$; $v_{k_x}=\frac{1}{\bar{h}}\frac{\partial \epsilon_k}{\partial k_x}$) caused by magnon accumulation, which results in an interconnected conductivity that arises from the interaction Hamiltonian stated above in eq.\ref{Hem}.\\

In order to comprehend the electronic current density, we must use the linearized Boltzmann equation to describe electronic transport with group velocity $v_{k_x}$ under an external electric field $\mathbf{E}$ along any desired direction $x$ as \cite{PhysRevB.96.024449, PhysRevB.105.184434, fert1969two}
 \begin{equation}
v_{k_x} \frac{\partial f_\sigma(x,\mathbf{k})}{\partial x} - e E_x v_{k_x} \frac{\partial f^0_\sigma}{\partial \epsilon_{\mathbf{k}}} = -\frac{f_\sigma(x,\mathbf{k}) - \langle f_\sigma(x) \rangle}{\tau_\sigma}- \frac{f_\sigma(x,\mathbf{k}) - \langle f_{-\sigma}(x) \rangle }{\tau_{\uparrow \downarrow}} 
+ \left[ \frac{\partial f_\sigma(x,\mathbf{k})}{\partial t} \right]_{em}
\end{equation}
where, $\tau_\sigma$ and $\tau_{\uparrow \downarrow}$ are the spin-conserving and spin-flipping electron relaxation time with spin $\sigma$, and  the last term is due to the interaction of electrons with magnons, which is important for us.\\

The last interaction terms in the above electron Boltzmann equations can be expressed  from the interaction Hamiltonian ($H_{em}$) as follows:\cite{PhysRevB.105.184434}
\begin{equation}
\left[ \frac{\partial f_\sigma(x,\mathbf{k})}{\partial t} \right]_{em} \sim -J_{em}^2S/N \sum_{\mathbf{q}} F(\mathbf{k}, \mathbf{k+q},\mathbf{q}, \mathbf{-q}, \uparrow, \downarrow);  \sigma = \uparrow, \downarrow
\end{equation}
where $F$ is a complex functional of $f_\sigma(x,\mathbf{k})$ and $g(x,\mathbf{q})$.
%As a result, the additional current could be expressed as $j_{em}=\sigma_{em}\frac{d}{dx}\delta n_m(x)$ in the extended Ohm's law \cite{PhysRevB.96.024449}, where interconnected conductivity induced by the exchange coupling is directly proportional to the $J_{em}^2$. This $J_{em}^2$ dependence is also same when a thermal gradient is applied across the magnetic insulator \cite{PhysRevB.86.214424}.\\
As a result, an additional current induced by the exchange coupling could be written as $j_{em} \sim \int d\mathbf{k} v_{k_x} \sum_{\sigma}\tau_\sigma\left[ \frac{\partial f_\sigma(x,\mathbf{k})}{\partial t} \right]_{em}$ \cite{PhysRevB.105.184434}, which  is directly proportional to the $J_{em}^2$. This $J_{em}^2$ dependence is also same when a thermal gradient is applied across the magnetic insulator \cite{PhysRevB.86.214424}.\\

}

\bibliography{cite}